\documentclass[prd,12pt,nofootinbib,tightenlines]{revtex4}

\usepackage{amssymb,caption,youngtab,graphicx,color}
\usepackage{amsmath,amsthm, amssymb}
\usepackage{xfrac,slashed,cancel}
\usepackage[smalltableaux]{ytableau}
\usepackage{simplewick}
\usepackage{enumerate}
\usepackage{multirow}
\usepackage{footnote}

\newcommand{\mb}{\mathbb}
\newcommand{\ba}{\begin{array}}
\newcommand{\ea}{\end{array}}
\newcommand{\bea}{\begin{eqnarray}}
\newcommand{\eea}{\end{eqnarray}}
\newcommand{\be}{\begin{equation}}
\newcommand{\ee}{\end{equation}}
\newcommand{\bpm}{\begin{pmatrix}}
\newcommand{\epm}{\end{pmatrix}}
\newcommand{\nn}{\nonumber}
\newcommand{\dst}{\displaystyle}

\newcommand{\qrq}{\quad\Rightarrow\quad}

\newcommand{\p}{\partial}
\newcommand{\al}{\alpha}
\newcommand{\da}{\delta}
\newcommand{\ta}{\theta}
\newcommand{\la}{\lambda}
\newcommand{\La}{\Lambda}
\def\a{\alpha}
\def\l{\lambda}
\def\g{\gamma}
\def\half{\frac{1}{2}}
\def\O{\Omega}
\newcommand{\si}{\sigma}
\newcommand{\ga}{\gamma}
\newcommand{\Ga}{\Gamma}
\newcommand{\om}{\omega}
\newcommand{\Om}{\Omega}
\newcommand{\ep}{\epsilon}

\newcommand{\yc}{\Yvcentermath1}
\newcommand{\curl}{{\rm curl}}

\newcommand{\slp}{{\slash\hspace{-5pt}\p}}  
\newcommand{\slx}{{\hspace{2pt}\slash\hspace{-6pt}x}} 
\newcommand{\slz}{{\hspace{2pt}\slash\hspace{-6pt}z}} 
\newcommand{\slD}{{\hspace{2pt}\slash\hspace{-7pt} D}}

\renewcommand{\phi}{\varphi}

\def\eq{\begin{equation}}
\def\eqe{\end{equation}}
\def\eqa{\begin{eqnarray}}
\def\eqae{\end{eqnarray}}

\newcommand\ytabl[1]{\ytableausetup{centertableaux}\begin{ytableau}#1\end{ytableau}}

\def\bop#1{\setbox0=\hbox{$#1M$}\mkern1.5mu
	\vbox{\hrule height0pt depth.04\ht0
	\hbox{\vrule width.04\ht0 height.9\ht0 \kern.9\ht0
	\vrule width.04\ht0}\hrule height.04\ht0}\mkern1.5mu}
\def\bo{{\mathpalette\bop{}}}
\newdimen\downsy	 \downsy=-1.5ex
\newcount\num
\def\boxes#1{
        \num=1
        \mskip-2.8mu
        \bo
        \loop
        \ifnum\num<#1
        \llap{\raise\num\downsy\hbox{$\bo$}}
        \advance\num by1
        \repeat}
\newdimen\upsy	 \upsy=.75ex
\newcount\numup
\def\boxup#1#2{
        \numup=#1
        \advance\numup by-1
        \mskip2.8mu
        \raise\numup\upsy\hbox{$#2$}}
\def\boxit#1{\leavevmode\thinspace\hbox{\vrule\vtop{\vbox{
	\hrule\kern1pt\hbox{\vphantom{\bf/}\thinspace{\bf#1}\thinspace}}
	\kern1pt\hrule}\vrule}\thinspace}

\def\hook#1{{\vrule height#1pt width0.4pt depth0pt}}
\def\leftrighthookfill#1{$\mathsurround=0pt \mathord\hook#1
        \hrulefill\mathord\hook#1$}
\def\underhook#1{\vtop{\ialign{##\crcr
        $\hfil\displaystyle{#1}\hfil$\crcr
        \noalign{\kern-1pt\nointerlineskip\vskip2pt}
        \leftrighthookfill5\crcr}}}
\def\thunderhook#1{\rlap{\kern.4em\underhook{\phantom{#1}}}{#1}}

\def\claw#1{{\vrule height0pt width0.4pt depth#1 pt}}
\def\leftrightclawfill#1{$\mathsurround=0pt \mathord\claw#1
        \hrulefill\mathord\claw#1$}
\def\overhook#1{\vbox{\ialign{##\crcr\noalign{\kern1pt}
       \leftrightclawfill5\crcr\noalign{\kern1pt\nointerlineskip}
       $\hfil\displaystyle{#1}\hfil$\crcr}}}

\def\upbracketfill{
       $\leaders\hrule\hfill$}
\def\underbracket#1{\mathop{\vtop{\ialign{##\crcr
   $\hfil\displaystyle{#1}\hfil$\crcr\noalign{\kern3pt\nointerlineskip}
   \upbracketfill\crcr\noalign{\kern3pt}}}}\limits}
\def\upslurfill{
       $\bracelu\leaders\vrule\hfill\braceru$}
\def\underslur#1{\mathop{\vtop{\ialign{##\crcr
   $\hfil\displaystyle{#1}\hfil$\crcr\noalign{\kern3pt\nointerlineskip}
   \upslurfill\crcr\noalign{\kern3pt}}}}\limits}

\begin{document}

\title{Spherical harmonics for the compactification of\\ IIB supergravity on $S^5$}
\label{pvn}

\author{P. van Nieuwenhuizen\footnote{Talk given on June 26, 2011 at the Max Kreuzer Memorial Conference in Vienna.}}
\affiliation{Stony Brook University, Stony Brook, NY 11794, U.S.A.}\email{vannieu@gmail.com}

\begin{abstract}
We give a detailed derivation of the spherical harmonics which correspond to the spectrum in the tables of the 1985 paper by Kim, Romans and the author on the Kaluza-Klein dimensional reduction of IIB supergravity on $S^5$ to 5-dimensional maximal anti de Sitter supergravity. We show how all methods that have been used in the literature to obtain spectra if the internal space is $S^n$ can be derived from embedding $S^n$ in $\mb{R}^{n+1}$. A crucial observation for spin 1/2 spherical harmonics is that one also needs a local Lorentz rotation if one transforms to polar or stereo\-graphic coordinates. The relation to the vector spherical harmo\-nics of electromagnetism is also worked out.
\end{abstract}
\maketitle

\section{Introduction}

In the 1980's many physicists considered the Kaluza-Klein (KK) compac\-tification of supergravity models and worked out their spectra (the values of the masses and their degeneracies). One such article \cite{KRN} dealt with the compactification of IIB supergravity on $S^5$, yielding maximal ($N=8$) supergravity in 5-dimensional anti de Sitter spacetime. The authors were  H. J. Kim who was a graduate student at Stony Brook, Larry Romans who was a graduate student at CalTech, and the author. We denote this 1985 paper by KRN in what follows. The article contains a series of tables and figures for the spectrum. Many people have used it for the AdS/CFT program of string theory, and one person (Ian Kogan) told me long ago that he had checked all formulas on the computer. However, the article is rather condensed and I have often been asked for more explanation and more background material. I always demurred because in KRN one particular method was used but I wanted first to understand how it was related to other methods which were also widely used by other physicists. In the course of teaching string courses at Stony Brook the answers came slowly to me, so that I now believe I can present a clear picture. The aim of this article is thus pedagogical: a discussion of the various methods to obtain spectra on $S^n$, and their interrelations. Experts in KK theory have each their own method, and may not even want to spend time on reading about other methods. 
We shall derive explicit expressions for the spherical harmonics (in particular Killing spinors) but such information is not necessary if one is only interested in the spectra.
In the conclusions we shall present a very quick method to obtain spectra on general coset manifolds $G/H$: all one needs to know are the Casimir operators of $G$ and $H$.

There exist excellent reviews on KK theory \cite{reviews}, where complete lists of references can be found. In this review we shall only refer to those articles that form the basis of our approach and with which we are thoroughly familiar.
Conference proceedings (as opposed to talks at conferences) should contain reviews that remain of interest as time goes on. Since Max Kreuzer also spent much time in updating his research in lectures, I decided to publish this material in these proceedings.

Let me now first write a few words about Max. Max Kreuzer carried String Theory at the Technical University (and, one might add, in all of Austria). Through Wolfgang Kummer, who died in 2007, I have been visiting the TU every J\"anner (J\"anner is January in Austrian) for the past 12 years, and I developed a friendship with Wolfgang, Toni Rebhan and Max. Every day we would have lunch together and drink coffee afterwards in a large office of graduate students. A few times we went to the famous TU ball in the imperial Hofburg, where Max would disappear in the ballroom for salsa dancing and be lost for hours. 
During my visits he got interested in work I had done with Antonio Grassi on the BRST quantization of the Berkovits superstring, and with Sebastian Guttenberg and Johanna Knapp they wrote an article (On the covariant quantization of type II superstring", hep-th/0405007). They did not solve the problem, nor did we, but their article is quite interesting. 

Twice Max helped me with solving a problem I had for a long time. One problem I encountered while writing a book on Advanced Quantum Gauge Field Theory. One can give a clear physical explanation of the asymptotic freedom of QCD if one works in the Coulomb gauge. However, calculations in the Coulomb gauge are tedious, even one-loop calculations. To my pleasant surprise I found an article by Max and Wolfgang which contained these calculations (Nucl.\ Phys.\ B {\bf 276} (1986) 466) but in very abridged form. Max went home, found his old calculations, and pieced together the explicit one-loop results I needed.

Another problem I ran into while teaching string theory. The spectrum of the closed bosonic (or spinning) string is defined by the BRST condition $(Q_L+Q_R)\left| phys\right>=0$, but the result is the direct product of $Q_L\left| phys\right>=0$ and $Q_R\left| phys\right>=0$. This fact is known as the K\"unneth formula in cohomology theory, but the proofs I found in math books were rather complicated. One day I explained to Max my frustration at finding a simple proof, and the next day he came back with an extremely simple proof, just based on linear algebra. I now use it in my string lectures.

Max taught superb classes on string theory whose (rather condensed) lecture notes one can find on the web \cite{kreuzer}. I followed several of his classes, and each time I felt sorry I had to leave by the end of J\"anner. With fond memory to Max and his classes, I dedicate this lecture to the good and interesting times we had together.

\section{Scalar spherical harmonics on $S^n$}

The construction of spherical harmonics for scalar fields on $S^n$ is well-known, but we present it here as an introduction to the more complicated cases of higher spin which we discuss later.
In the Kaluza-Klein dimensional reduction, a scalar field $B(x,z)$ in $D$-dimensional Minkowski space with coordinates $(x,z)$ is expanded in terms of harmonics on a compact space with coordinates $z$ in the following manner
\bea
B(x,z)=\sum_{I_1}B_{I_1}(x) Y^{I_1}(z)
\label{eq1}
\eea
The $Y^{I_1}(z)$ are eigenfunctions of the Laplacian on $S^n$, which we shall denote by $\Box_S$, and $B_{I_1}(x)$ are fields in $d$-dimensional (usually anti de Sitter) space with coordinates $x$. We shall no longer consider these fields $B_{I_1}$ and coordinates $x$ in what follows, but concentrate on $Y^{I_1}(z)$. The symbol $I_1$ indicates that these fields have one component on $S^n$. For example, vector harmonics on $S^5$ are denoted in KRN by $Y_\alpha^{I_5}$ if they are transversal $(D^\alpha Y_\alpha^{I_5}=0)$ and by $D_\alpha Y^{I_1}$ if they are longitudinal (not transversal). Here $D_\alpha$ is the covariant derivative on $S^n$. 

A word of caution. In KRN the $D$-dimensional field equations are reduced on $S^5$ to field equations of the form 
\bea
(\Box_x+\text{more})\varphi(x)Y(z)=(\Box_z+\text{more})\varphi(x)Y(z)
\eea
where the ``$+\text{more}$" on the left-hand side is part of the field equations in $d$ dimensions, and the ``$+\text{more}$" on the right-hand side is part of the mass operator. (Actually, mixing occurs, but never more than $2\times2$ mixing, and one can rather easily diagonalize the field equations.) The parts ``$+\text{more}$" are easy to evaluate because the Riemann and Ricci tensors can be expressed in terms of the metric (as always for a maximally symmetric space). The hard problem is the determination of the eigenvalues and their degeneracies of $\Box_z$. This is the problem discussed in this article. Due to the ``$+\text{more}$" in the mass operator, the figures in KRN are shifted with respect to the figures we construct below, but the degeneracies are the same. Also, in KRN the mass spectrum of all scalars in $d$ dimensions is plotted in one figure, but these scalars correspond to scalar, vector, and tensor harmonics.

What are the $Y^{I_1}(z)$ in (\ref{eq1})? The $Y_{l m}(\theta,\varphi)$ of quantum mechanics suggest an answer.
\bea
\left.
\ba{l}
Y_{0,0}=1; \quad Y_{1,\pm1}\sim\sin\theta e^{\pm i\varphi}=x\pm i y\\
Y_{1,0}\sim\cos\theta=z; \quad
Y_{2,0}\sim 3\cos^2\theta-1=2z^2-x^2-y^2 \\
Y_{2,\pm1}\sim(x\pm i y)z; \quad
Y_{2,\pm2}\sim(x+i y)^2
\ea\right\} \quad \ba{c} \text{all at}\\ r=1 \ea
\label{eq3}
\eea
These are all homogeneous traceless polynomials in $x$, $y$, $z$ at radius $r=1$. We therefore begin with a homogeneous polynomial in $\bar x^\mu$ where $\bar x^\mu$ ($\mu=1,\dots,n+1$) are the Cartesian coordinates of the flat embedding space $\mb{R}^{n+1}$. We can formally write these polynomials as \cite{chodos,rubin}
\bea
\bar P^{(l)}(\bar x)=c_{\mu_1\dots\mu_l}\bar x^{\mu_1}\dots\bar x^{\mu_l}
\eea
We require that $\bar\Box\bar P=0$ where $\bar\Box=\frac{\p}{\p\bar x^\mu}\frac{\p}{\p\bar x^\nu}\da^{\mu\nu}$. Then $c$ is traceless, $\da^{\mu\nu}c_{\mu\nu\mu_3\dots\mu_l}=0$ (recall $Y_{2,0}\sim 2z^2-x^2-y^2$). We go over to polar coordinates $\hat y^\mu=(r,\ta^\al)$ where $r$ is radius of $S^n$ and $\ta^\al$ ($\al=1,\dots,n$) are any angles on $S^n$. The caret $\,\hat{}\,$ will always denote quantities in the system with coordinates $\hat y$. The metric in polar coordinates is of the form
\bea
\hat g_{\mu\nu}(\hat y)=\left(\ba{c@{\quad}c} 1 & 0 \\ 0 & r^2 g_{\al\beta}(\ta) \ea\right)
\label{metric-polar}
\eea
Then $\hat g=\det\hat g_{\mu\nu}=r^{2n}g(\ta)$, and 
\bea
\bar\Box=\hat\Box=\frac{1}{\sqrt{\hat g}}\frac{\p}{\p\hat y^\mu}\sqrt{\hat g}\hat g^{\mu\nu}\frac{\p}{\p\hat y^\nu}
=\frac{1}{r^n}\p_r r^n\p_r+\frac{1}{r^2}\Box_S(\ta)
\eea
where $\Box_S(\ta)$ is the Laplacian on $S^n$. Since $\hat P^{(l)}(\bar x)$ is homogeneous in $\bar x$, it factorizes 
\bea
\bar P^{(l)}(\bar x)=\hat P^{(l)}(\hat y)=r^l Y(\ta)
\eea
and substituting this expression into $\bar\Box\bar P^{(l)}=0$ one arrives at
\bea
\bar\Box\bar P^{(l)}(\bar x)=\hat\Box\hat P^{(l)}(\hat y)
=\Big[\frac{1}{r^2}l(l+n-1)+\frac{1}{r^2}\Box_S(\ta)\Big]r^l Y(\ta)=0
\eea
Hence the eigenvalues $\la_s$ of $-\Box_S(\ta)$ on $S^n$ are given by
\bea
\la_s(n,l)=l(n+l-1); \quad l=0,1,2,\dots \quad (s \text{ for scalar})
\eea

On $S^2$ with the usual $Y_{l m}$ one finds the familiar result $\la_s=l(l+1)$ of quantum mechanics, and on $S^1$ one finds $\la_s=l^2$, the 1926 result of Klein and Fock. On $S^5$ we find $\la(5,l)=l(l+4)$.

To determine the degeneracy $d_s(n,l)$ of these eigenvalues, we note \cite{chodos,rubin} that it is given by the number of polynomials $\bar P^{(l)}(\bar x)$, that is, it is equal to the number of symmetric polynomials of order $l$ in $n+1$ dimensions, minus the number of such polynomials of order $l-2$ (the trace)
\bea
d_s(n,l)=\left(\ba{c} n+l\\ l\ea\right)-\left(\ba{c}n+l-2\\l-2\ea\right)
\eea
The index $I_1$ in (\ref{eq1}) labels these harmonics, see (\ref{eq3}).
On $S^2$ one finds $d_s=2l+1$, and on $S^1$ one finds $d_s=2$ except for the massless mode with $l=0$ which is not degenerate. On $S^5$, the case of interest, we find the following spectrum
\begin{center} 
\includegraphics[clip=true]{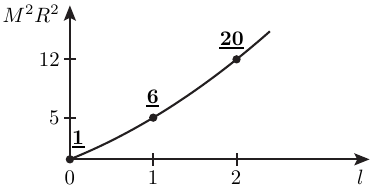}
\centerline{Fig.~1. $S^5$, scalars (cf. Figure 2 of KRN)}
\end{center}
There is one harmonic with vanishing eigenvalue, the constant, then 6 harmonics $\bar x^\mu$, and 20 harmonics $\bar x^\mu\bar x^\nu-\frac16\da^{\mu\nu}\bar x^2$, etc. In Figure 2 of KRN one finds several curves for different $x$-space scalars whose masses are due to different harmonics (scalar, vector, tensor harmonics), but one of these curves corresponds to scalar harmonics and agrees with our Figure~1.

Another popular set of coordinates on $S^n$ are the stereographic coordinates $\hat y^\nu=(z^\al,R)$ with $\al=1,\dots,n$ and $R$ the radius of $S^n$ (we write $R$ instead of $r$ to indicate that we are using stereographic coordinates). The metric in stereographic coordinates is not diagonal
\bea
\hat g^{\mu\nu}=\left(\ba{c@{\qquad}c} 
\dst\left(\frac{4R^2+z^2}{4R^2}\right)^2\da^{\al\beta}+\frac{z^\al z^\beta}{R^2} & z^\al/R \\[12pt]
z^\beta/R & 1 \ea\right)
\eea
as follows from squaring the vielbein field we obtain in the section on spin 1/2 spherical harmonics (see equation (\ref{eq59})). We discuss spherical harmonics in stereographic coordinates further at the end of that section. They give, of course, the same spectrum as polar coordinates.

\section{Vector spherical harmonics}

The transversal vector spherical harmonics $Y_\al(\ta)$ on $S^n$ satisfying $D^\al(\ta)Y_\al(\ta)=0$ are obtained \cite{chodos,rubin} from vector fields $\bar P_\mu^{(l)}(\bar x)$ whose components $\mu=1,\dots, n+1$ are homogeneous polynomials in $\bar x^\mu$. We impose the equation $\bar\Box\bar P_\mu^{(l)}=0$ which implies that each component is traceless.  We choose a basis in which only one component of $\bar P_\mu^{(l)}$ is nonzero, and we call this the $\mu$-th component. For example
\bea
\bpm 0\\0\\2z^2-x^2-y^2\epm \quad\text{has}\quad \ba{c} n=2\\l=2\\\mu=3\ea 
\eea
and is an example of $\bar P_\mu^{(l)}(\bar x)$.

We go over to polar coordinates $\hat y^\nu=(r,\ta^\al)$ with $\al=1,\dots,n$, and by substitution and the chain rule we obtain
\bea
\bar\Box\bar P_\mu^{(l)}=\frac{\p\hat y^\nu}{\p\bar x^\mu}\hat\Box\hat P_\nu^{(l)}=0 \qrq \hat\Box\hat P_r^{(l)}=0 \quad\text{and}\quad \hat\Box\hat P_\al^{(l)}=0
\eea
where
\bea
\hat P_r^{(l)}(\hat y) &=& \frac{\p\bar x^\mu}{\p r}\bar P_\mu^{(l)}(\bar x)=\frac{\bar x^\mu}{r}\bar P_\mu^{(l)}(\bar x)=r^l\rho(\ta) \nn\\
\hat P_\al^{(l)} &=& \frac{\p\bar x^\mu}{\p\ta^\al}\bar P_\mu^{(l)}(\bar x)=r^{l+1}V_\al(\ta)
\eea
Here $\hat\Box$ is the Laplacian in $n+1$ dimensions in polar coordinates, which is of course different for $\hat P_r^{(l)}$ and $\hat P_\al^{(l)}$. We decompose $V_\al$ into a longitudinal and a transversal part
\bea
V_\al(\ta)=\p_\al\si(\ta)+Y_\al(\ta); \quad D^\al(\ta)Y_\al(\ta)=0
\label{decomp-V}
\eea
One can obtain $\si$ by solving $D^\al\p_\al\si=D^\al V_\al$ and then $Y_\al$ are given by $V_\al-\p_\al\si$. Often one can directly check that in a given expression for $\p_\al\si+Y_\al$ the $Y_\al$ are transversal. The $Y_\al(\ta)$ are the vector spherical harmonics we are interested in.

Neither $\rho$ nor $\si$ is a scalar spherical harmonic because neither is totally symmetric or traceless (as we shall check in examples), but linear combinations of $\rho$ and $\si$ yield the scalar spherical harmonics $Y^{l+1}(\ta)$ and $Y^{(l-1)}(\ta)$. A Young tableau illustrates these decompositions \cite{chodos,rubin}
\bea
\ba{cccccccccc}
\yc\yng(1) & \otimes & \yc\yng(2) &=& \yc\yng(3) & \oplus & \yc\yng(2,1) & \oplus & \yc\yng(1) & \quad\text{(for $l=2$)} \\
\dst\frac{\p\bar x^\mu}{\p\hat y^\nu} &\times & \bar P_\mu^{(l)} &=& \rho,\si &+& Y_\al &+& \rho,\si &
\ea
\label{eq14}
\eea
On the left-hand side the first box corresponds to the index $\mu$ of $\bar{x}^\mu$, and it takes into account that we are considering vectors. It is multiplied by the Young tableau of the symmetric traceless polynomial which one finds in one entry of the vector. The total number of terms on the left-hand side is $(n+1)$ for $\bar{x}^\mu$ times $d_s(n,l)$ for the number of scalar spherical harmonics in one entry of the vector.

The degeneracy of the eigenvalues of the $Y_\al$ is clear from this picture: it is equal to the number of terms on the left-hand side minus the number of the two scalar spherical harmonics
\bea
d_v(n,l)=(n+1)d_s(n,l)-d_s(n,l+1)-d_s(n,l-1)
\eea

To obtain the eigenvalues $\la_v(n,l)$ defined by $-\Box_S Y_\al=\la_v(n,l)Y_\al$ we must take the transversal part of $\hat\Box\hat P_\al^{(l)}=0$. There are three sets of terms in $\hat\Box\hat P_\al^{(l)}$: purely radial terms with $\p/\p r$ and powers of $r$, terms which constitute $\frac{1}{r^2}\Box_S(\ta)$ for a vector, and mixed terms due to the Christoffel symbols $\hat\Ga_{\beta\ga}^r$ and $\hat\Ga_{r\al}^\beta$. The nonvanishing connections for the metric in (\ref{metric-polar}) are
\bea
\hat\Ga_{r\al}^\beta=\frac{1}{r}\da_\al^\beta; \quad \hat\Ga_{\al\beta}^r=-r g_{\al\beta}(\ta) \quad\text{and}\quad \hat\Ga_{\al\beta}^\ga=\Ga_{\al\beta}^\ga(\ta)
\eea
After having decomposed $\hat\Box\hat P_\al^{(l)}$ into the three sets of terms, one may use the following identities to project out the transversal part
\bea
D^\al(\ta)\Box_S(\ta)Y_\al(\ta)=0; \quad \Box_S\p_\al\si=\p_\al\Box_S\si-R_{\al\beta}(\ta)D^\beta\si
\eea
There is still some tedious (but perfectly straightforward) algebra involved, but the result is very simple
\bea
\la_v(n,l)=l(n+l-1)-1
\eea

Let us give some examples.

{\bf\boldmath $l=0$}.
For $l=0$ the polynomial $\bar P_\mu^{(0)}$ is a constant, and $V_\al=\p_\al\bar x^\mu$. Hence for $l=0$ there is only a longitudinal part, but no $Y_\al$.

{\bf\boldmath $l=1$}.
$\bar P_\rho^{(1)}=\bar x^\nu$ if $\rho=\mu$, and zero otherwise. Then $\hat P_r^{(1)}=\frac{\p\bar x^\mu}{\p r}\bar x^\nu=\frac{\bar x^\mu\bar x^\nu}{r}$ and $\hat P_\al^{(1)}=\frac{\p\bar x^\mu}{\p\ta^\al}\bar x^\nu$. The decomposition of $\hat P_\al^{(1)}=r^2 V_\al$ into a total derivative and a transversal part according to (\ref{decomp-V}) is as follows
\bea
\hat P_\al^{(1)}(r,\ta)=\frac{\p}{\p\ta^\al}\left(\frac12\bar x^\mu\bar x^\nu\right)+\frac12\left(\frac{\p\bar x^\mu}{\p\ta^\al}\bar x^\nu-\frac{\p\bar x^\nu}{\p\ta^\al}\bar x^\mu\right) \nn
\eea
so
\bea
\rho(\ta) &=& \frac{\bar x^\mu\bar x^\nu}{r^2};\quad \si=\frac12\frac{\bar x^\mu\bar x^\nu}{r^2}+c\da^{\mu\nu}; \nn\\
Y_\al(\ta) &=& \frac{1}{2r^2}(\p_\al\bar x^\mu \bar x^\nu-\p_\al\bar x^\nu\bar x^\mu) \qquad
\eea
where $c$ is an integration constant.
(To prove that $D^\al Y_\al=0$, note that $D^\al\p_\al\bar x^\rho$ is proportional to $\bar x^\rho$ because $\bar x^\rho$ is a scalar spherical harmonic, the harmonic with $l=1$.) Hence the two scalar harmonics that are linear combinations of $\rho$ and $\si$, and the transversal vector harmonic $Y_\al$ are given by
\bea
\yc\yng(2) &=& \left(1+\frac{1}{2c}\frac{1}{n+1}\right)\rho+\left(-\frac{1}{c}\frac{1}{n+1}\right)\si = \frac{\bar x^\mu \bar x^\nu}{r^2} - \frac{1}{n+1}\delta^{\mu\nu} \nn\\[5pt]
\bullet &=& \frac{1}{c}\left(\si-\frac12\rho\right) = \delta^{\mu\nu}; \quad
Y_\al=\yc\young(\nu,\mu)=\frac{1}{2r^2}(\p_\al\bar x^\mu \bar x^\nu-\p_\al\bar x^\nu\bar x^\mu)
\label{eq22}
\eea
These $Y_\al$ are the Killing vectors on $S^n$, of which there are $\frac12(n+1)n$ (namely $\frac12 n(n-1)$ rotations and $n$ translations). They satisfy the Killing equation $D_\al K_\beta+D_\beta K_\al=0$, as is clear by substituting (\ref{eq22}) and using $D_\al \partial_\beta \bar x^\mu = \frac{1}{n}g_{\al \beta}\Box \bar x^\mu$ (the latter is proven by showing that the norm of $D_\al\partial_\beta\bar x^\mu - \frac{1}{n}g_{\al\beta}\Box\bar x^\mu$ vanishes). Acting with $D^\al$ on this equation one confirms that $\la_v(n,l=1)=n-1$.~\footnote{
Use that $R_{\al\beta\ga\da}(\Ga(\ta))=-g_{\al\ga}g_{\beta\da}+g_{\al\da}g_{\beta\ga}$ on $S^n$, hence $R_{\al\ga}\equiv R_{\al\beta\ga}{}^\beta=-(n-1)g_{\al\ga}$. (In our conventions, the scalar curvature $R=R_{\al\beta}g^{\al\beta}$ of $S^n$ is negative.)
}

{\bf\boldmath $l=2$}.
This is fun. We choose $\bar P_\si^{(2)}=\bar x^\nu\bar x^\rho-\frac{1}{n+1}\da^{\nu\rho}\bar x^2$ for $\si=\mu$. Then, defining $\bar P^{\nu\rho}=\bar x^\nu\bar x^\rho-\frac{1}{n+1}\da^{\nu\rho}\bar x^2$ we obtain
\bea
\rho(\ta)=\frac{\bar x^\mu}{r^3}\bar P^{\nu\rho}; \quad V_\al(\ta)=\frac{1}{r^3}\frac{\p\bar x^\mu}{\p\ta^\al}\bar P^{\nu\rho}
\eea
The decomposition of $V_\al$ into a total derivative and a transversal part requires more work
\bea
r^3\si(\ta) &=& \frac{1}{3}\bar x^\mu\bar x^\nu\bar x^\rho
-\frac{1}{3n}(\da^{\mu\nu}\bar x^\rho+\da^{\mu\rho}\bar x^\nu)\bar x^2
+\left(\frac{2}{3n}
-\frac{1}{n+1}\right)\da^{\nu\rho}\bar x^\mu\bar x^2 \nn\\
r^3 Y_\al(\ta) &=& \yc\young(\nu\rho,\mu)=\frac{2}{3}\p_\al\bar x^\mu\bar Q^{\nu\rho}-\frac{1}{3}\p_\al\bar x^\nu\bar Q^{\mu\rho}-\frac{1}{3}\p_\al\bar x^\rho\bar Q^{\mu\nu}
\label{eqA}
\eea
where $\bar Q^{\mu\nu}=\bar x^\mu \bar x^\nu-\frac{1}{n}\da^{\mu\nu}\bar x^2$.
Note that there is no integration constant in $\si$ possible because $\si$ has an odd number (3) of indices. To prove that this is the correct decomposition, note that $\p_\al\si+Y_\al=V_\al$ and $D^\al Y_\al=0$.~\footnote{
Use $D^\al\p_\al\bar x^\mu=-n\bar x^\mu$, and $\frac{1}{r^2}(\p_\al\bar x^\mu)D^\al\bar x^\nu=\da^{\mu\nu}-\frac{1}{r^2}\bar x^\mu\bar x^\nu$ (because $\frac{1}{r^2}\p_\al\bar x^\mu D^\al\bar x^\nu=\da^{\rho\si}\p_\rho\bar x^\mu\p_\si\bar x^\nu-\p_r\bar x^\mu\p_r\bar x^\nu$).
}
Note that $Y_\al$ has the symmetries of the Young tableau: first antisymmetrize in $\mu\nu$, then symmetrize in $\nu\rho$. It is also traceless in all 3 pairs of indices. Referring to (\ref{eq14}) we find
\bea
\yc\yng(3) \sim \rho+n\si &=& \frac{n+3}{3}\left[\bar x^\mu\bar x^\nu\bar x^\rho-\frac{1}{n+3}(\da^{\mu\nu}\bar x^\rho+\da^{\mu\rho}\bar x^\nu+\da^{\nu\rho}\bar x^\mu)\bar x^2\right] \nn\\
\yc\yng(1) \sim \rho-3\si &=& \frac1n(\da^{\mu\nu}\bar x^\rho+\da^{\mu\rho}\bar x^\nu)\bar x^2-3\left(\frac{2}{3n}-\frac{1}{n+1}\right)\da^{\nu\rho}\bar x^\mu\bar x^2
\eea

{\bf\boldmath $l=3$}.
Finally we do the case of $l = 3$. We start from $\bar P_\si^{(3)} = \delta_{\si \mu}\bar P^{(3)}$; $\bar P^{(3)} = \bar x^\nu \bar x^\rho \bar x^\si - \frac{1}{n+3}(\da^{\mu\nu}\bar x^\rho + \da^{\mu\rho} \bar x^{\nu} + \da^{\nu\rho} \bar x^\mu)$ and find
\bea
   V_{\alpha}(\theta) &=& \partial_\al \bar x^\mu \bar P^{(3)}
\eea
The quickest way to obtain $Y_\al$ is to assume that it has the Young symmetry $Y_\al = \yc\young(\nu\rho\sigma,\mu)$ and to start from
\bea
   Y_\al &=& \frac{3}{4}\partial_\al \bar x^\mu \bar x^\nu \bar x^\rho \bar x^\si - \frac{1}{4}\bar x^\mu(\partial_\al \bar x^\nu \bar x^\rho \bar x^\si + \bar x^\nu \partial_\alpha \bar x^\rho \bar x^\si + \bar x^\nu \bar x^\rho \partial_\alpha \bar x^\sigma) \text{ minus traces}
\eea
and to fix the trace terms by imposing
\bea
   \da_{\mu\nu} Y_{\al}^{\mu; \nu\rho\si} &=& 0; \qquad \da_{\nu\rho}Y_{\al}^{\mu; \nu\rho\si} = 0
\eea
The trace terms can only have the form
\bea
   \text{traces} &=& a\,\partial_\al \bar x^\mu(\da^{\nu\rho}\bar x^\si + \da^{\nu\si}\bar x^\rho + \da^{\rho\si}\bar x^\nu) \nn\\[5pt]
   && +b\,(\da^{\mu\nu}\partial_\al \bar x^\rho x^\si + \text{5 terms due to symmetry in $\nu\rho\si$)}\bar x^2\nn\\[5pt]
   && +c\,\bar x^\mu(\da^{\nu\rho}\partial_\al \bar x^\si + \da^{\nu\si}\bar x^\rho + \da^{\rho\si} \bar x^\nu)
\eea
One finds
\bea
  a &=& \frac{1}{n+3}\left(\frac{1}{4}\frac{n-1}{n+1} - 1\right); \qquad b = \frac{1}{4(n+1)}; \qquad c = \frac{1}{4}\frac{n-1}{n+1}\frac{1}{n+3}
\eea
A more constructive way, though also more laborious, is to first compute $D^\al V_\al = D^\al \partial_\al \si$, next to solve for $\si$, then to evaluate $D_\al \si$, and finally to construct $V_\al - \partial_\al \si = Y_\al$. The expression for $\si$ that one finds in this way is as follows
\bea
  \si &=& \frac{1}{4}\bar x^\mu \bar x^\nu \bar x^\rho \bar x^\si - \frac{1}{4(n+1)}(\da^{\mu\nu}\bar x^\rho \bar x^\si + \da^{\mu\rho}\bar x^\nu \bar x^\si + \da^{\mu\si}\bar x^\nu \bar x^\rho)\bar x^2\nn\\[5pt]
      && +\,\al\,(\da^{\mu\nu}\da^{\rho\si} + \da^{\mu\rho}\da^{\nu\si} + \da^{\mu\si}\da^{\nu\rho})\bar x^4 \nn\\[5pt]
      && -\,\frac{1}{4}\frac{n-1}{n+1}\frac{1}{n+3}\bar x^\mu (\da^{\nu\rho}\bar x^\si + \da^{\nu\si}\bar x^\rho + \da^{\rho\si} \bar x^\nu)\bar x^2
\eea
The coefficient $\al$ cannot be fixed in this way because $\partial_\al \bar x^4$ = 0, but it can be fixed by requiring that linear combinations of the polynomials $\rho$ and $\si$ form the Young tableaux
\bea
 p + q &\sim& \yc\yng(4) \qquad \text{and}\qquad r + s \sim \yc\yng(2) \nn
\eea

We can now exhibit the spectrum of transversal vector fields on $S^5$ in a similar figure as for the scalars
\begin{center}
\includegraphics[clip=true]{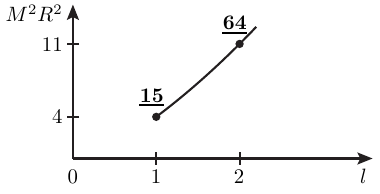}
\centerline{Fig.~2. $S^5$, vectors (cf. Figure 1 of KRN)}
\end{center}
The number of Killing vectors on $S^5$ is 15, namely 5 for translations and 10 for rotations.

\subsection{The vector spherical harmonics in electromagnetism.}

In Jackson's book on Electromagnetism \cite{Jackson} one can find ``multipole expansions" of the electric and magnetic fields $\vec E$ and $\vec B$ with time dependence $e^{-i\om t}$ in a source-free region. These fields are decomposed into radial parts and angular parts, and we want to clarify in this section the relation of these angular parts to the vector spherical harmonics $Y_\al(\ta)$ satisfying $D^\al Y_\al=0$ which we discussed before. There are differences
\begin{itemize}
\item[(i)] the wave equations for the electric and magnetic fields are $(\nabla^2+k^2)\vec E=0$ and $(\nabla^2+k^2)\vec B=0$ where $k=\om/c$, instead of $\nabla^2\vec E=\nabla^2\vec B=0$;
\item[(ii)] the Maxwell equations impose constraints on $\vec E$ and $\vec B$: ${\rm div}\vec E={\rm div}\vec B=0$ and 
$\curl\,\vec E=i k\vec B$, $\curl\,\vec B=-i k\vec E$.
\end{itemize}
However, one would expect that the angular parts of vectors on $S^2$ in both approaches are related. 

The multipole expansion of $\vec E$ is given by \cite{Jackson}
\bea
\vec E &=& \sum_{l,m}\Bigg[a_M(l,m)g_l(k r)\frac{1}{\sqrt{l(l+1)}}\vec L Y_{l m}(\ta,\varphi) \nn\\
&&\hspace{40pt}
+\frac{i}{k}a_E(l,m)\,\curl\left(f_l(k r)\frac{1}{\sqrt{l(l+1)}}\vec L Y_{l m}(\ta,\varphi)\right)\Bigg]
\eea
and a similar expression for $\vec B$. The functions $f_l(k r)$ and $g_l(k r)$ are spherical Bessel functions, $a_E(l,m)$ and $a_M(l,m)$ are arbitrary constants, and $\vec L=-i\vec r\times\vec\nabla$. Clearly, these vectors are given in Cartesian coordinates.

The first term in $\vec E$ is transversal due to $\vec r\cdot\vec L=0$, but the second term contains both a longitudinal part (in the near zone $\vec E\rightarrow -\vec\nabla Y_{l m}/r^{l+1}$)  and a transversal part (in the radiation zone both $\vec E$ and $\vec B$ become transversal). The same holds for $\vec B$. Thus the $\vec L Y_{l m}$, when written in polar coordinates, are equal to our $V_\al(r,\ta)$. Furthermore, since $\vec\nabla\cdot\vec L=0$, these $V_\al(r,\ta)$ satisfy $D^\al V_\al=0$, and hence {\bf\boldmath the $\vec L Y_{l m}$ are the transversal vector harmonics $Y_\al(\ta)$ we have constructed before, but in Cartesian coordinates}.

Let us check these statements in a few examples.

{\bf\boldmath $Y_{1m}$}. Taking $Y_{l m}$ to be given by $x/r$, we find
\bea
\vec L\;\frac{x}{r}=\bpm 0 \\ z/r \\ -y/r \epm
\eea
and in polar coordinates $(r,\ta^\al)$ with $\ta^\al=(\ta,\varphi)$
\bea
\left(\vec L\;\frac{x}{r}\right)_r\equiv\frac{\vec r}{r}\cdot\vec L\;\frac{x}{r}=0, \quad
\left(\vec L\;\frac{x}{r}\right)_\al=\frac{\p y}{\p\ta^\al}\frac{z}{r}-\frac{\p z}{\p\ta^\al}\frac{y}{r}
\eea
The vectors $\frac{1}{r^2}\left(\frac{\p y}{\p\ta^\al} z-\frac{\p z}{\p\ta^\al} y\right)$ are Killing vectors on $S^2$ as we have shown previously, hence they are indeed the $Y_\al$ which are obtained from $\bar P_\mu(\bar x)=\bar x^\nu$.

{\bf\boldmath $Y_{2m}$}. Next consider $Y_{l m}=x y/r^2$. Then one finds
\bea
\vec L\,Y_{l m} &=& \frac{1}{r^2}\bpm -x z\\ y z\\ x^2-y^2 \epm, \quad
(\vec L\,Y_{l m})_r=0 \nn\\
(\vec L\,Y_{l m})_\al &=& \frac{\p x}{\p\ta^\al}(-x z)+\frac{\p y}{\p\ta^\al}(y z)+\frac{\p z}{\p\ta^\al}(x^2-y^2)
\label{eqE}
\eea
In order to compare with the $Y_\al$ we obtained from polynomials $\bar P_\mu(\bar x)$, we consider in turn the polynomials $-x z$, then $y z$, and finally $x^2-\frac{1}{n}\vec r\,{}^2=\frac12(x^2-y^2-z^2)$ and $-(y^2-\frac{1}{n}\vec r\,{}^2)=-\frac12(y^2-x^2-z^2)$. According to (\ref{eqA}) they give the following vector harmonics $Y_\al$
\bea
-x z &:& \frac23\frac{\p x}{\p\ta^\al}(-x z)-\frac13\frac{\p x}{\p\ta^\al}(-x z)-\frac13\frac{\p z}{\p\ta^\al}\left(-\frac12 x^2+\frac12 y^2+\frac12 z^2\right) \nn\\
y z &:& \frac23\frac{\p y}{\p\ta^\al}(y z)-\frac13\frac{\p y}{\p z}(y z)+\frac13\frac{\p z}{\p\ta^\al}\left(\frac12 y^2-\frac12 x^2-\frac12 z^2\right) \nn
\eea\bea
\frac12(x^2-y^2-z^2) &:& \frac23\frac{\p z}{\p\ta^\al}\left(\frac12 x^2-\frac12 y^2-\frac12 z^2\right)-\frac13\frac{\p x}{\p\ta^\al}(x z)-\frac13\frac{\p x}{\p\ta^\al}(x z) \nn\\
-\frac12(y^2-x^2-z^2) &:& \frac23\frac{\p z}{\p\ta^\al}\left(-\frac12 y^2+\frac12 x^2+\frac12 z^2\right)-\frac13\frac{\p y}{\p\ta^\al}(y z)-\frac13\frac{\p y}{\p\ta^\al}(y z) \nn\\
\eea
The sum of these terms is indeed equal to (\ref{eqE}).

Let us next consider the terms in $\vec E$ of the form $\curl\,\vec L(f_l(k r)Y_{l m})$. It is useful to decompose the operator $\curl\,\vec L$ as follows \cite{Jackson}
\bea
i\vec\nabla\times\vec L=\ep^{i j k}\nabla_j(\ep_{k l m}x^l\nabla^m)=\vec r\,\nabla^2-\vec\nabla\left(1+r\frac{\p}{\p r}\right)
\eea 
The functions $f_l(k r)Y_{l m}$ satisfy the wave equation $(\nabla^2+k^2)f_l(k r)Y_{l m}=0$, hence the second term in the multipole expansion of $\vec E$ reads
\bea
-\frac{1}{k}a_E(l,m)\left[k^2\vec r+\vec\nabla\left(1+r\frac{\p}{\p r}\right)\right]\frac{g_l(k r)}{\sqrt{l(l+1)}}Y_{l m}(\ta,\varphi)
\eea
The first term contributes only to $\rho$ (it lies along $\vec r$), but the second term is a gradient, hence it has a part with $\vec\nabla=\frac{\p}{\p r}$ which contributes to $\rho$,~\footnote{%
This part can be simplified using 
$$
\frac{\p}{\p r}\left(1+r\frac{\p}{\p r}\right)g_l(k r)=\left(-k^2+\frac{l(l+1)}{r}\right)g_l(k r) $$
} 
and a part due to $\nabla=\frac{\p}{\p\ta^\al}$ which contributes to $\si$.

Hence the vector spherical harmonics of electromagnetism coincide with the harmonics we have discussed. In particular $\vec L\,Y_{l m}$ yields only $Y_\al$ but no $\p_\al\si$. Note that these simple formulas only make sense on $S^2$.

\section{Spin 1/2 spherical harmonics}

The explicit construction of harmonics for scalars and vectors was discussed in references {\,}\cite{chodos} and {\,}\cite{rubin}{\,}, but the case of spinorial harmonics was not tackled there, and turned out to be more complicated. A short conversation with Gary Gibbons at a conference in India was helpful. He had determined the spectrum of spin 1/2 harmonics on $S^3$ by choosing the natural diagonal vielbein field in polar coordinates, and then writing down the Dirac equation  \cite{gibbons}. However, the form of the spinor harmonics is not determined in this way (nor was it needed for his purposes). Several discussions over the years with Martin Rocek helped me further. Eventually it became clear to me that one should introduce in addition to a change of coordinates also a local Lorentz rotation, and the problem was to determine the relation between this local Lorentz rotation and the vielbein field in the new coordinates.

We start \cite{gibbons} with the Dirac equation for a massless spin 1/2 field in the embedding flat $(n+1)$-dimensional Euclidean space with Cartesian coordinates~\footnote{
Our conventions are that all $\ga^m$ are hermitian and $\ga^m\ga^m=I$ (no summation over $m$). In $n+1$ dimensions they are $2^{[\frac{n+1}{2}]}\times2^{[\frac{n+1}{2}]}$ matrices.
}
\bea
\bar\slp\bar\psi(\bar x)\equiv\ga^m\da_m^\mu\frac{\p}{\p\bar x^\mu}\bar\psi^A(\bar x)=0
\eea
where $A$ is the spinor index and the matrices $\ga^m$ are constant Dirac matrices in $n+1$ dimensions. The field $\bar\psi$ is a spinor (nonchiral if $n+1$ is even) that is homogeneous of order $l$ in $\bar x^\mu$. We construct a basis such that $\bar\psi(\bar x)=\bar x^{\mu_1}\dots\bar x^{\mu_l}\al+\text{``trace terms"}$, where the constant spinor $\al$ has only one nonvanishing component, and the terms denoted by ``trace terms" are terms depending on $\bar\slx=\bar x^\mu\da_\mu^m\ga_m$ such that the Dirac equation holds.

For example, for $l=1$ we get 
\bea
\bar\psi^{(1)}(\bar x)=\bar x^\mu\al-\frac{1}{n+1}\bar\slx\ga^\mu\al
\eea
It is clear that $\bar\slp\bar\psi^{(1)}=0$ since $\bar\slp\bar\slx=n+1$. If $n+1=3$ there are 6 such spinors for $l=1$, but the Dirac equation yields 2 constraints, hence there are 4 linearly independent $\bar\psi^{(1)}(\bar x)$ on $S^2$. 

For $l=2$ we obtain
\bea
\bar\psi^{(2)}(\bar x)=\left[\bar x^\mu \bar x^\nu-\frac{1}{n+3}(\bar\slx\ga^\mu x^\nu+\bar\slx\ga^\nu\bar x^\mu+\bar x^2\da^{\mu\nu})\right]\al
\eea
It is not only ``gamma traceless" (due to the Dirac equation, see the next paragraph) but also traceless in the ordinary sense (as is obvious by taking the trace over $\mu$ and $\nu$), as it should be since $\bar\slp\bar\slp\bar\psi(\bar x)=\bar\Box\bar\psi(\bar x)$.

One can formally write the spinor $\bar\psi(\bar x)$ as 
\bea
\bar\psi(\bar x)=\bar x^{\mu_1}\dots\bar x^{\mu_l}c^A{}_{\mu_1\dots\mu_l}
\eea
where $A$ is the spinor index. Clearly $c$ is symmetric in $\mu_1\dots\mu_l$ and gamma-traceless
\bea
(\ga^{\mu_1})^A{}_B c^B{}_{\mu_1\mu_2\dots\mu_l}=0
\eea
Contracting with $\ga^{\mu_2}$ shows that it is also traceless in the ordinary sense
\bea
\da^{\mu\nu}c^A{}_{\mu\nu\mu_3\dots\mu_l}=0
\eea
We shall not use this formulation in terms of $c$.

The number of linearly independent spinors $\bar\psi^{(l)}(\bar x)$ is equal to the number of components of $\al$ times the number of polynomials $\bar x_{\mu_1}\dots\bar x_{\mu_l}$ in $n+1$ dimensions minus the number of polynomials with one index less (to account for the Dirac equation)
\bea
\text{number of }\bar\psi^{(l)}(\bar x)=\left[\bpm n+l\\ l\epm -\bpm n+l-1\\l-1 \epm\right]2^{[\frac{n+1}{2}]}
\eea

Next we transform to coordinates $\hat y^\nu$. In practice these new coordinates are either polar coordinates $(r,\ta^\al)$ or stereographic coordinates $(z^\al,R)$, but for the time being we leave the choice open and proceed formally. The Dirac equation becomes
\bea
\ga^m\left(\da_m^\mu\frac{\p\hat y^\nu}{\p\bar x^\mu}\right)\frac{\p}{\p\hat y^\nu}\hat\psi(\hat y)=0; \quad \hat\psi(\hat y)=\bar\psi(\bar x)
\eea
The composite object $\da_m^\mu\frac{\p\hat y^\nu}{\p\bar x^\mu}$ is the vielbein field $E^\nu{}_m(\hat y)$ in $n+1$ dimensions, but it is non-diagonal because the tangent frames are still along the axes of the $\bar x$ system. When we choose polar coordinates there is a natural vielbein field that is diagonal \cite{gibbons} (see below), so this requires to rotate the tangent frames such that their axes become aligned along the polar coordinate axes. This motivates us to apply also in the general case a local Lorentz transformation with Lorentz matrix $\La(\hat y)$ in the spinor representation~\footnote{
In the case of stereographic coordinates we shall first choose a suitable coset representative in the vector representation for the coset $S^n=SO(n+1)/SO(n)$. The Cartan-Maurer equations will then give us the corresponding vielbein field and from it we shall then determine the matrix $\La$ for this case.
}
($\La$ is reducible when $n+1$ is even; we discuss the consequences later).

To implement a local Lorentz transformation by substitution we insert unity in front of $\hat\psi(\hat y)$, as $\hat\psi=\La\La^{-1}\hat\psi$, and define
\bea
\psi(\hat y)=\La^{-1}(\hat y)\hat\psi(\hat y)
\label{eq41}
\eea
The Dirac equation for $\psi(\hat y)$ can then be written as follows after adding a factor $\La^{-1}$ in front of the equation
\bea
(\La^{-1}\ga^m\La)\left(\da_m^\mu\frac{\p\hat y^\nu}{\p\bar x^\mu}\right)\left(\frac{\p}{\p\hat y^\nu}+(\La^{-1}\frac{\p}{\p\hat y^\nu}\La)\right)\psi(\hat y)=0
\eea
If the finite local Lorentz transformation is given by the $(n+1)\times(n+1)$ matrix $L^m{}_n=(\exp\la)^m{}_n$, where $\la^m{}_n$ lies in the algebra of $SO(n+1)$, then $\La=\exp\frac14\la^{m n}\ga_m\ga_n$. It follows that 
\bea
(\La^{-1}\ga^m\La)=L^m{}_n\ga^n
\eea
and substituting this back into the Dirac equation yields
\bea
\ga^n\left[\frac{\p\hat y^\nu}{\p\bar x^\mu}\da_m^\mu L^m{}_n\right]\left\{\frac{\p}{\p\hat y^\nu}+(\La^{-1}\frac{\p}{\p\hat y^\nu}\La)\right\}\psi(\hat y)=0
\eea
The vielbein field is now given by
\bea
E^\nu{}_n=\frac{\p\hat y^\nu}{\p\bar x^\mu}\da_m^\mu L^m{}_n
\label{eqV}
\eea
Given $E^\nu{}_n$ one can determine $L^m{}_n$ (as in the case of polar coordinates), or given $L^m{}_n$ one can calculate the corresponding $E^\nu{}_n$ (as in the case of stereographic coordinates).

The operator $\frac{\p}{\p\hat y^\nu}+(\La^{-1}\frac{\p}{\p\hat y^\nu}\La)$ is the covariant derivative for spin 1/2 spinors in $(n+1)$ dimensions. The object $(\La^{-1}\frac{\p}{\p\hat y^\nu}\La)$ is the spin connection in frames that are rotated with respect to the $\bar x$ frames; it is, of course, pure gauge because the spin connection in the $\bar x$ system vanishes. There is another way of looking at this result which will be useful when we discuss the approach with stereographic coordinates. Invert (\ref{eqV}) to obtain $E^n{}_\nu=(L^{-1})^n{}_m\da^m_\mu\frac{\p\bar x^\mu}{\p\hat y^\nu}$ and multiply by $d\hat y^\nu$. This yields the one-form equation
\bea
E^n=(L^{-1})^n{}_m d\bar x^m \quad\text{with}\quad E^n=E^n{}_\nu d\hat y^\nu
\eea
The spin connection $\Omega^{m n}$ which one obtains from the Cartan-Maurer equation $d E^m+\Om^m{}_n\wedge E^n=0$ is then $\Om=L^{-1}d L$, and this agrees with $\La^{-1} d\La=\frac14(L^{-1}d L)^{m n}\ga_m\ga_n$.

The Dirac equation, still in $n+1$ dimensions and in general coordinates, can be succinctly written as
\bea
\hat\slD\psi=0, \quad \hat\slD=E^\nu{}_m\ga^m\left(\frac{\p}{\p\hat y^\nu}+\frac14\Omega_\nu{}^{m n}\ga_m\ga_n\right)
\label{Dirac-eq}
\eea
Let us now first specialize to polar coordinates, and then to stereographic coordinates.

\subsection{Polar coordinates}

For definiteness we consider the case $n=3$, although all formulas hold for general $n$. We set
\bea
(d s)^2 &=& d\bar x^2+d\bar y^2+d\bar z^2+d\bar v^2 \nn\\
&&\bar x=r\cos\ta, \quad
\bar y=r\sin\ta\cos\varphi \nn\\
&&\bar z=r\sin\ta\sin\varphi\cos\chi, \quad
\bar v=r\sin\ta\sin\varphi\sin\chi \nn\\
(d s)^2 &=& d r^2+r^2 d\ta^2+r^2\sin^2\ta d\varphi^2+r^2\sin^2\ta\sin^2\varphi d\chi^2
\eea
A natural set of diagonal vielbein one-forms is then \cite{gibbons}
\bea
E^1=d r, \quad E^2=r d\ta, \quad E^3=r\sin\ta d\varphi, \quad E^4=r\sin\ta\sin\varphi d\chi
\eea
The corresponding spin connections are obtained by solving the Cartan-Maurer equation $d E^m+\Om^{m n}\wedge E^n=0$. The solution has the following nonvanishing components \cite{gibbons}
\bea
&& \Om^{21}=d\ta, \quad \Om^{31}=\sin\ta d\varphi, \quad \Om^{32}=\cos\ta d\varphi \nn\\
&& \Om^{41}=\sin\ta\sin\varphi d\chi, \quad \Om^{42}=\cos\ta\sin\varphi d\chi, \quad \Om^{43}=\sin\ta\cos\varphi d\chi \qquad
\eea
Actually, all we need in $n+1$ dimensions is that $\Om_r{}^{p q}=0$ and that $\Om^{a1}=\frac{1}{r}E^a{}_\al d\ta^\al$. This is easily proven using that $E^1=d r$ and $E^{a}$ for $a\neq 1$ is proportional to $r$.

For $n=3$ the Dirac equation for $\psi(\hat y)=\La^{-1}\hat\psi(\hat y)$ in polar coordinates and in tangent frames aligned along the coordinate axes reads
\bea
&&\Big[\ga^1\p_r+\frac14\frac{\ga^2}{r}(2\ga^2\ga^1)+\frac14\frac{\ga^3}{r\sin\ta}(2\sin\ta\ga^3\ga^1) \nn\\
&&\quad +\frac{\ga^4}{4r\sin\ta\sin\varphi}(2\sin\ta\sin\varphi\ga^4\ga^1)+\frac{1}{r}\slD_S(\ta)\Big]\psi(\hat y)=0
\eea
We shall determine the matrix $\La$ later, see equation (\ref{eq77}).
The operator $\slD_S(\ta)$ is the Dirac operator on $S^n$; it depends only on the angles $\ta$. Note that the second, third and fourth terms are equal. Generalizing to $n$ dimensions yields 
\bea
-\ga^1\left(\p_r+\frac{n}{2}\frac{1}{r}\right)\psi(\hat y)=\frac{1}{r}\slD_S(\ta)\psi(\hat y)
\label{eq49}
\eea
Since $\psi=\La^{-1}\hat\psi=\La^{-1}\bar\psi(\bar x)$ is still homogeneous in $r$ (because $\La$ does not depend on $r$), we have
\bea
\psi(r,\ta)=r^l\psi(\ta)
\eea
To obtain the eigenvalues of $\slD_S$ we must remove the matrix $\ga^1$ in the Dirac equation in (\ref{eq49}). This is achieved in different ways for $n+1$ even and $n+1$ odd.

{\bf\boldmath Even $n+1$}: we choose a suitable representation
\bea
\ga^1=\bpm 0 & I \\ I & 0 \epm, \quad \ga^{a+1}=\bpm 0 & -i\si^a \\ i\si^a & 0 \epm \quad\text{for }a=1,\dots,n
\eea
where $\si^a$ is any set of Dirac matrices in $n$ dimensions. Setting $\psi=\bpm \psi_{+} \\ \psi_{-} \epm$, we obtain two diagonal equations
\bea
\left.
\ba{lcr}
-i\slD_S\psi_{-} &=& -(l+\frac{n}{2})\psi_{-}\\[5pt]
+i\slD_S\psi_{+} &=& -(l+\frac{n}{2})\psi_{+}
\ea
\right\} \qquad
\ba{c}
\la_{spinor}(n,l)=\pm(l+\frac{n}{2}) \\[5pt]
l=0,1,2,\dots
\ea
\eea

{\bf\boldmath Odd $n+1$}: we contract with $1\pm i\ga^1$ (which are \emph{not} projection operators) using $(1\pm i\ga^1)\ga^1=\pm i(1\mp i\ga^1)$. Then we get \cite{gibbons}
\bea
\slD_S[(1\mp i\ga^1)\psi]=\mp i \left(l+\frac{n}{2}\right)[(1\mp i\ga^1)\psi]
\eea
so the same eigenvalues.

The degeneracies of these eigenvalues are the number of spinors $\bar\psi^{(l)}$ (or half as many if $n+1$ is even). Thus the spectrum is
\bea
\la_{spinor}(n,l) &=& \pm\left(l+\frac{n}{2}\right) \nn\\
d_{spinor}(n,l) &=& \left[\bpm n+l \\ l \epm - \bpm n+l-1 \\ l-1 \epm\right]2^{[\frac{n}{2}]}
\label{eq54}
\eea
For $S^5$ we find the following spectrum (eigenvalues $\slD_S\hat\psi=i\la_{spinor}(n,l)\hat\psi$ and degeneracy)
\begin{center}
\includegraphics[clip=true]{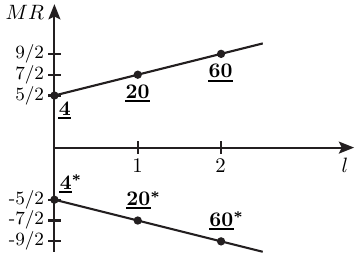}
\centerline{Fig.~3. $S^5$, spin 1/2 (cf. Figure 5 of KRN)} 
\end{center}

\subsection{Stereographic coordinates}

Stereographic coordinates $\hat y^\nu=(z^\al,R)$ with $\al=1,\dots,n$ and $R$ again the radius are related to the Cartesian coordinates $\bar x^\mu$ with $\mu=1,\dots,n+1$ by projecting onto a plane. Using that $\frac{\bar x^\al}{z^\al}=\frac{\bar x^{n+1}+R}{2R}$ (see the figure) one finds
\bea
\ba[b]{c}
\includegraphics[clip=true,scale=1]{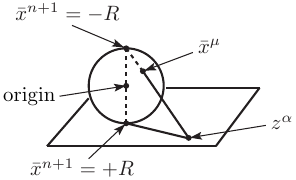}
\ea
\quad
\ba[b]{l}
\dst \bar x^\al=\frac{z^\al 4 R^2}{4R^2+z^2}, \quad \bar x^{n+1}=R\frac{4R^2-z^2}{4R^2+z^2} \\[15pt]
\dst \da^{\mu\nu}\bar x^\mu\bar x^\nu=R^2, \quad z^2\equiv z^\al z^\beta\da^{\al\beta} \\[5pt]
{}
\ea
\eea
Note that $\bar x^{n+1}$ at the South Pole is positive. As coset representative $L^m{}_n$ for the coset $S^n=SO(n+1)/SO(n)$ we take an $SO(n+1)$ matrix which maps the South Pole (the point with $\bar x^{n+1}=R$) to the point $\bar x^\mu$
\bea
L^m{}_n=\left(\ba{c@{\qquad}c} \dst\da^a{}_b-\frac{2z^a z^b}{4R^2+z^2} &\dst \frac{4R z^a}{4R^2+z^2} \\[10pt] 
\dst \frac{-4 R z^b}{4R^2+z^2} & \dst\frac{4R^2-z^2}{4R^2+z^2} \ea\right)
\eea
(For $z\rightarrow 0$ we obtain the unit matrix, and for this reason we choose $x^{n+1}=+R$ instead of $-R$ for the South Pole).
We use these matrices for the local Lorentz rotation which acts on the tangent frames. We find then from the relation between vielbein fields and local Lorentz rotations in (\ref{eqV})
\bea
L^m{}_n E^n{}_\nu=\da^m{}_\mu\frac{\p\bar x^\mu}{\p\hat y^\nu}
\eea
One can solve for $E^n{}_\nu$ (a rather tedious calculation but the result is easy to check) to obtain
\bea
E^n{}_\nu=\bpm \dst\frac{4 R^2}{4R^2+z^2}\da^a{}_\al \quad & \dst\frac{-4z^a R}{4R^2+z^2} \\[10pt] 0 & 1 \epm
\eea
where $a=1,\dots,n$ is a flat vector index on $S^n$ and $\al=1,\dots,n$ is a curved vector index on $S^n$.
The result $E^a{}_\al=\da^a{}_\al(4R^2)/(4R^2+z^2)$ is well-known,~\footnote{
In the literature one usually sets $R=1/2$ in which case $E^a{}_\al=\da^a{}_\al/(1+z^2)$. We keep the dependence on $R$ because we shall have to differentiate with respect to $R$ later.
}
but there is an off-diagonal term $E^a{}_{n+1}=-4z^a R/(4R^2+z^2)$. We need the inverse vielbein for the Dirac equation. It is given by
\bea
\label{eq59}
E^\nu{}_n=\bpm \dst \frac{4R^2+z^2}{4R^2}\da^\al{}_a \quad & \dst \frac{z^\al}{R} \\[10pt] 0 & 1 \epm
\eea
The Dirac equation in (\ref{Dirac-eq}) becomes
\bea
E^\al{}_a\ga^a\left[\frac{\p}{\p z^\al}+(\La^{-1}\frac{\p}{\p z^\al}\La)\right]\psi(z,R) \nn\\
+\ga^{n+1}\frac{z^\al}{R}\left[\frac{\p}{\p z^\al}+(\La^{-1}\frac{\p}{\p z^\al}\La)\right]\psi(z,R) \nn\\
+\ga^{n+1}\left[\frac{\p}{\p R}+(\La^{-1}\frac{\p}{\p R}\La)\right]\psi(z,R) &=& 0
\eea
The connections are given by
\bea
(L^{-1}\frac{\p}{\p z^\al}L)^{a b}=\frac{1}{4R^2+z^2}
\bpm 
-2\da^a{}_\al z^b+2\da^b{}_\al z^a \quad & 4 R \da^a{}_\al \\[5pt]
-4 R\da^b{}_\al & 0
\epm
\eea
Then~\footnote{
From $L^{-1}d L=e^a K_a^{(v)}+\frac12\om^{a b}H_{a b}^{(v)}$ and
\bea
K_a^{(v)}=\bpm &&&&0\\&&&&\vdots\\&&&&2\\&&&&\vdots\\0&\dots&-2&\dots&0\epm, \quad
H_{a b}^{(v)}=\bpm &&&&0\\
&&&&\\
&&\ba{ccc} &&1\\&&\\-1&&\ea&&\vdots\\
&&&&\\
0&&\dotfill&&0\epm
\eea
we obtain $[K_a,K_b]=-4H_{a b}$. Then in the (reducible if $n+1=\text{even}$) spinor representation we get
$K_a^{(s)}=\ga_a\ga_{n+1}$ and $H_{a b}^{(s)}=\frac12\ga_{a b}$.
}
\bea
(\La^{-1}\frac{\p}{\p z^\al}\La) &=& e^a{}_\al\ga_a\ga_{n+1}+\frac14\om_\al{}^{a b}\ga_a\ga_b \nn\\
&& e^a{}_\al=\frac{2 R\da^a{}_\al}{4R^2+z^2}, \quad
\om_\al{}^{a b}=\frac{-2\da^a{}_\al z^b+2\da^b{}_\al z^a}{4R^2+z^2} 
\eea
Similarly
\bea
L^{-1}\frac{\p}{\p R}L &=& \bpm 0 & -4z^a/(4R^2+z^2) \\[5pt] 4z^b/(4R^2+z^2) & 0 \epm \nn\\[5pt]
\La^{-1}\frac{\p}{\p R}\La &=& \frac{-2z^a}{4R^2+z^2}\ga_a\ga_{n+1}=\frac{-2\slz}{4R^2+z^2}\ga_{n+1}
\eea
Substituting these results into the Dirac equation yields
\bea
E^\al{}_a\ga^a\Big[D_\al\psi+e^b{}_\al\ga_b\ga_{n+1}\Big]\psi
+\ga^{n+1}\frac{z^\al}{R}\left[\frac{\p}{\p z^\al}+e^a{}_\al\ga_a\ga_{n+1}\right]\psi \nn\\
+\ga^{n+1}\left[\frac{\p}{\p R}-\frac{2\slz}{4R^2+z^2}\ga_{n+1}\right]\psi &=& 0 \qquad
\eea
where we used that $z^\al\om_\al{}^{a b}=0$ to replace $z^\al D_\al$ in the second line by $z^\al\p_\al$. The last term in the first line cancels the last term in the second line.
The dimensionless $E^\al{}_a$ is part of the $(n+1)$-dimensional vielbein field, and the dimensionful $e^\al{}_a$ is the vielbein on the coset $S^n$. They are related by
\bea
E^\al{}_a=\frac{1}{2R}e^\al{}_a \quad\text{where}\quad e^\al{}_a e^a{}_\beta=\da^\al{}_\beta
\eea
Using this, the Dirac equation reduces to
\bea
\frac{1}{2R}e^\al{}_a\ga^a\Big[D_\al^{(S)}+e^b{}_\al\ga_b\ga_{n+1}\Big]\psi(z,R)
=-\frac{\ga^{n+1}}{R}\left[z^\al\frac{\p}{\p z^\al}+R\frac{\p}{\p R}\right]\psi(z,R) \nn\\
\eea
The operator $z^\al\frac{\p}{\p z^\al}+R\frac{\p}{\p R}$ counts the number of $\bar x^\mu$ in the homogeneous polynomial $\bar\psi(\bar x)$, hence
\bea
\slD_S\psi(z,R)=-\frac{\ga^{n+1}}{R}\left[l+\frac{n}{2}\right]\psi(z,R)
\eea
where $\slD_S=E^\al{}_a\ga^a\left(\frac{\p}{\p z^\al}+\dots\right)$ has dimension $(\text{length})^{-1}$. From here on we remove the matrix $\ga^{n+1}$ the same way as we removed $\ga^1$ in the case of polar coordinates. Hence the spectrum in stereographic coordinates is the same as in polar coordinates, as it should of course be the case.

\section{The KRN method}

In KRN the spin 1/2 spherical harmonics $\Xi^{(l)\pm}(z)$ were constructed as a product of scalar spherical harmonics $Y^{(l)}(z)$ and Killing vectors $\eta^{\pm}(z)$,
\bea
\Xi^{(l)\pm}=A_{\pm}Y^l(\ta)\eta^{\pm}(\ta)+B_{\pm}(\slD_S Y^l)\eta^{\pm}(\ta)
\label{eqP}
\eea
The Killing spinors satisfy the equation $D_\al\eta^{\pm}(z)=\pm\frac{i}{2}c\tau_\al\eta^{\pm}$, with $c=1/R$, and requiring that this ansatz satisfies $i\slD_S\Xi^{(l)\pm}=\la_{\pm}\Xi^{(l)\pm}$ yields the equations
\bea
A_{\pm}\pm\frac{1}{2}c B_{\pm}(n-2)=\la_{\pm}B_{\pm} \nn\\
A_{\pm}\left(\mp\frac{1}{2}n c\right) + B_{\pm}c^2 l(l+n-1)=\la_{\pm}A_{\pm}
\eea
where we used $-\Box_S Y^l=c^2 l(l+n-1)Y^l(\ta)$. Requiring that the determinant of the matrix of this system vanishes, one finds a quadratic equation for $\la_{\pm}$, and two solutions for $\la_{+}$, and two for $\la_{-}$
\bea
\la_{+}=-\frac{c}{2}\pm c\left(l+\frac{n-1}{2}\right)=\left\{\ba{l} +c\left(l+\frac{n-2}{2}\right) \\[3pt] -c\left(l+\frac{n}{2}\right) \ea\right. \nn\\
\la_{-}=+\frac{c}{2}\pm c\left(l+\frac{n-1}{2}\right)=\left\{\ba{l} +c\left(l+\frac{n}{2}\right) \\[3pt] -c\left(l+\frac{n-2}{2}\right) \ea\right.
\eea
The corresponding eigenspinors on $S^5$ are given by
\bea
\left.
\ba{l}
\slD\Big[ c l Y^l\eta^{+}-i(\slD Y^l)\eta^{+}\Big]=i c\left(l+\frac32\right)\Big[\text{same}\Big] \\[5pt]
\slD\Big[ c (l+4) Y^l\eta^{+}+i(\slD Y^l)\eta^{+}\Big]=-i c\left(l+\frac52\right)\Big[\text{same}\Big]
\ea
\right\} \quad \text{for $\la_{-}$} 
\nn\\
\left.
\ba{l}
\slD\Big[ c (l+4) Y^l\eta^{-}-i(\slD Y^l)\eta^{-}\Big]=i c\left(l+\frac52\right)\Big[\text{same}\Big] \\[5pt]
\slD\Big[ c l Y^l\eta^{-}+i(\slD Y^l)\eta^{-}\Big]=-i c\left(l+\frac32\right)\Big[\text{same}\Big]
\ea
\right\} \quad \text{for $\la_{+}$}
\eea
The eigenfunctions corresponding to $l+3/2$ for $l=0$ clearly vanish, hence all eigenvalues are given by $\la(n,l)=\pm(l+n/2)$ with $l=0,1,2,\dots$ For $l=0$ there are two eigenfunctions with the eigenvalue $\la=n/2=5/2$
\bea
\slD\Big[Y^1\eta^{-}-i(\slD Y^1)\eta^{-}\Big] &=& i c\frac52\Big[Y^1\eta^{-}-i(\slD Y^1)\eta^{-}\Big] \nn\\
\slD\Big[4Y^0\eta^{+}\Big] &=& i c\frac52\Big[4Y^0\eta^{+}\Big]
\eea
The first set seems to constrain 6 Killing spinors since $Y^1 = (x, y, z)$, but it yields the same set of Killing spinors as the second set. For example one finds by direct evaluations
\bea
  Y^1 \eta^{-}_{\rm I} - i(\slD Y^1)\eta^{-}_{\rm I} &=& \left\{
                                                     \ba{cl}
                                                         \eta^{+}_{\rm I} & \text{for } Y^1 = z\\
                                                         -i\eta^{+}_{\rm II} & \text{for } Y^1 = x\\
                                                         \eta^{+}_{\rm II} & \text{for } Y^1 = y
                                                     \ea
                                                 \right.
\eea
The degeneracy of the $l=0$ spinor harmonics $\eta^{+}(\ta)$ and $\eta^{-}(\ta)$ on $S^5$ is 4 since one can choose $\eta^{\pm}$ at $\ta=0$ at will. This agrees with (\ref{eq54}). For $l=1$ the degeneracy seems to be $4\times6$ because in $Y^1\eta^{+}$ there are six ways to choose $Y^1$ and four for $\eta^{+}$. However, the Dirac equation subtracts one spinor, hence the degeneracy for $l=1$ is $4\times5$, again in agreement with (\ref{eq54}). We have now a good idea about the structure of the spinorial harmonics but we still have no explicit expressions for $\eta^{\pm}$.

At this point a puzzle arises: in the expressions we obtained in (\ref{eq41}) for the spin 1/2 spherical harmonics we did not find derivatives of $Y^l$, but in (\ref{eqP}) they are needed to find solutions to the eigenvalue equations. How can these two expressions be equal?

To solve this problem we return to the embedding method. 
To obtain explicit expressions for the spinor harmonics themselves we need to determine the matrix $\La$. We consider first polar coordinates; later we discuss the stereographic coordinates. Recall the relation $E^\nu{}_n=\frac{\p\hat y^\nu}{\p\bar x^\mu}\da^\mu{}_m L^m{}_n$. From it one can solve for $L^m{}_n$, namely
\bea
L^m{}_n=\da^m{}_\mu\frac{\p\bar x^\mu}{\p\hat y^\nu}E^\nu{}_n
\label{eq77}
\eea
Using the diagonal vielbein of the previous section, the result for $L^m{}_n$ on $S^3$ is
\bea
L^m{}_n = \left( \begin{array}{c@{\;\;\;\;}c@{\;\;\;\;}c@{\;\;\;\;}cc} 
\cos \theta & - \sin \theta  & 0 & 0 \\
 \sin \theta \cos \varphi & \cos \theta \cos \varphi & - \sin \varphi & 0 \\
\sin \theta \sin \varphi \cos \chi  & \cos \theta  \sin \varphi \cos \chi & \cos \varphi \cos \chi & - \sin \chi \\
\sin \theta \sin \varphi \sin \chi  & \cos \theta \sin \varphi \sin \chi  & \cos \varphi \sin \chi & \cos \chi \end{array} \right)
\label{eq304}
\eea
It can be written as the product of a rotation in the $x y$ plane, followed by a rotation in the $y z$ plane, etc.
\bea
L^m{}_n  &=&   \left( \begin{array}{ccccc}  1 & 0 & 0 & 0 \\  0 & 1 &  0 &   0 \\  0 & 0 & \cos \chi  & - \sin \chi  \\ 0 & 0 & \sin \chi  & \cos \chi \end{array} \right) 
  \left( \begin{array}{ccccc}  1 & 0 & 0 & 0 \\  0 & \cos \varphi &  - \sin \varphi &   0 \\ 0 &  \sin \varphi  & \cos \varphi  & 0 \\ 0 & 0 & 0 & 1 \end{array} \right)   \left( \begin{array}{ccccc}  \cos \theta & - \sin \theta & 0 & 0 \\  \sin \theta & \cos \theta & 0 & 0 \\ 0 & 0 & 1 & 0 \\ 0 & 0 & 0 & 1 \end{array} \right)  \nn\\
&&\hspace{-20pt} =  (\exp \chi L_{34}) (\exp \varphi L_{23} ) ( \exp \theta L_{12} )  \; \mbox{with} \; L_{ij} = \left( \begin{array}{cc} 0 & - 1 \\  1 & 0  \end{array} \right) \; \mbox{for} \; i < j  
\label{eq305}
\eea
As one may check, the first rotation aligns the $x$-axis along the radius, the second rotation aligns the $y$-axis in the $\ta$-direction, etc. Hence the matrix $\La^{-1}$ in (\ref{eq41}) is given by
\bea
\La^{-1}=e^{\frac12\ta\ga^1\ga^2}e^{\frac12\varphi\ga^2\ga^3}e^{\frac12\chi\ga^3\ga^4}
\eea
The spinor harmonics are then given in $n+1=$odd dimensions by multiplication by $(1\pm i\ga^1)$, and in $n+1=$even dimensions by taking the chiral and antichiral parts.

Let us determine the first few spinor harmonics explicitly.

{\bf\boldmath $l=0$}. The $l=0$ case yields Killing spinors. One starts from $\bar\psi=\hat\psi=\al=$constant, and obtains $\psi=\La^{-1}\al$. On $S^2$ we have $\La^{-1}=\exp\frac12\ta\ga^1\ga^2\exp\frac12\varphi\ga^2\ga^3$. Choosing $\ga^1=\si^3$, $\ga^2=\si^1$ and $\ga^3=\si^2$ where $\si^j$ are the Pauli matrices, we get 4 Killing spinors after multiplication by $(1\pm i\ga^1)$
\bea
\eta_{\rm I}^{+}=e^{i\varphi/2}\bpm \cos\frac{\ta}{2}(1+i) \\[3pt] -\sin\frac{\ta}{2}(1-i) \epm, \quad
\eta_{\rm I}^{-}=e^{i\varphi/2}\bpm \cos\frac{\ta}{2}(1-i) \\[3pt] -\sin\frac{\ta}{2}(1+i) \epm \nn\\[5pt]
\eta_{\rm II}^{+}=e^{-i\varphi/2}\bpm \sin\frac{\ta}{2}(1+i) \\[3pt] \cos\frac{\ta}{2}(1-i) \epm, \quad
\eta_{\rm II}^{-}=e^{-i\varphi/2}\bpm \sin\frac{\ta}{2}(1-i) \\[3pt] \cos\frac{\ta}{2}(1+i) \epm
\eea
They can be rewritten in easier form as
\bea
\eta_{\rm I}^{+}=e^{i\varphi/2}\bpm \cos\frac{\ta}{2} \\[3pt] i\sin\frac{\ta}{2} \epm, \quad
\eta_{\rm II}^{+}=e^{-i\varphi/2}\bpm i\sin\frac{\ta}{2} \\[3pt] \cos\frac{\ta}{2} \epm \nn\\[5pt]
\eta_{\rm I}^{-}=e^{i\varphi/2}\bpm \cos\frac{\ta}{2} \\[3pt] -i\sin\frac{\ta}{2} \epm, \quad
\eta_{\rm II}^{-}=e^{-i\varphi/2}\bpm -i\sin\frac{\ta}{2} \\[3pt] \cos\frac{\ta}{2} \epm
\eea
One may check that they satisfy the Killing spinor equations $D_\al\eta^{\pm}=\pm\frac{i}{2}e^a{}_\al\ga_a\eta^{\pm}$, which read in more explicit form
\bea
\p_\ta\eta^{\pm}=\pm\frac{i}{2}\si_1\eta^{\pm}, \quad
\left(\p_\varphi\mp\frac{i}{2}\cos\ta \,\si^3\right)\eta^{\pm}=\pm\frac{i}{2}\sin\ta\,\si^2\eta^{\pm}
\eea
They produce the 3 Killing vectors on $S^2$ for rotations around the $x, y$ or $z$ axis as follows
\bea
  \eta^{+\dagger}_{\rm I} \tau_\al \eta^{+}_{\rm I} = -\eta^{+\dagger}_{\rm II}\tau_\al \eta^{+}_{\rm II} = k_{(z)\al} = \bpm 0\\\sin^2\theta\epm; \quad \eta^{+\dagger}_{I}\tau_\al \eta^{+}_{\rm II} = k_{(y)\al} + i k_{(x)\al}\nn\\[5pt]
  k_{(x)} = \bpm-\sin\varphi\\ -\cos\varphi\sin\theta\cos\theta\epm; \quad k_{(y)} = \bpm\cos\varphi\\-\sin\varphi\sin\theta\cos\theta\epm
\eea
The Killing spinors $\eta^{-}_{\rm I}$ and $\eta^{-}_{\rm II}$ produce the same Kiling vectors
\bea
   \eta^{-\dagger}_{\rm I} \tau_\al \eta^{-}_{I} = -\eta^{-\dagger}_{\rm II}\tau_\al \eta_{\rm II}^{-} = -k_{(z)\al}; \quad \eta^{-\dagger}_{I}\tau_\al \eta_{\rm II}^{-} = k_{(y)\al} + i k_{(x)\al}
\eea
{\bf\boldmath $l=1$}. For $l=1$ we have $\bar\psi^{(l=1)}(\bar x)=\left[\bar x^\mu-\frac{1}{n+1}\bar x^\nu\ga^\nu\ga^\mu\right]\al$ with constant $\ga^\nu$, $\ga^\mu$ and $\al$. We find then for $\psi=\La^{-1}\hat\psi$
\bea
\psi^{(l=1)}=\bar x^\mu\La^{-1}\al-\frac{1}{n+1}\bar x^\nu(\La^{-1}\ga^\nu\La)(\La^{-1}\ga^\mu\La)(\La^{-1}\al)
\eea
The factor $\bar x^\nu\La^{-1}\ga^\nu\La=\bar x^\nu\da_{\nu n}L^n{}_m\ga^m$ is only nonvanishing if $m=1$, see (\ref{eq304}), hence this factor yields $r\ga^1$. The factor $\La^{-1}\ga^\mu\La$ can also be simplified if one uses $\La^{-1}\ga^\mu\La=L^\mu{}_n\ga^n$ and $L^\mu{}_n=\frac{\p\bar x^\mu}{\p\hat y^\nu}E^\nu{}_n\ga^n$. One finds then
\bea
\La^{-1}\ga^\mu\La=\frac{\p\bar x^\mu}{\p\hat y^\nu}E^\nu{}_n\ga^n
=\frac{\bar x^\mu}{r}\ga^1+\frac{1}{r}\slD_S\bar x^\mu
\eea
where we define $\slD_S\bar x^\mu=\frac{\p\bar x^\mu}{\p\ta^\al}\frac{1}{r}\ga^a e^\al{}_a(\ta)$. Hence
\bea
\psi^{(l=1)}=\left[\bar x^\mu-\frac{1}{n+1}r\ga^1\left(\frac{\bar x^\mu}{r}\ga^1+\frac{1}{r}\slD_S(\ta)\bar x^\mu\right)\right]\al
\eea
After multiplication by $(1\pm i\ga^1)$ we arrive at 
\bea
\Xi^{l=1,\pm} &=& \left[\frac{n}{n+1}\bar x^\mu\mp\frac{i}{n+1}(\slD_S\bar x^\mu)\right](1\pm i)\al \nn\\
&=& \frac{1}{n+1}\Big[n Y^1\mp i(\slD_S Y^1)\Big]\eta^{\pm}
\eea
where $Y^1=\bar x^\mu$. This agrees with KRN, eqs. (3.9) and (3.10) (except for the $+$ sign in (3.10) which should clearly be a $-$ sign).

So this explains the terms with $(\slD Y^l)$ in the ansatz of KRN: they are due to the Lorentz matrix $L$ in $\La^{-1}\ga^m\La$.

\section{Conclusion}

We have shown how to obtain the spectra, but also the spherical harmonics themselves, for spin 0, 1, 1/2 on $S^n$ by starting in a flat Euclidean $(n+1)$-dimensional embedding space with Cartesian coordinates $\bar x^\mu$, and then transforming to general coordinates on $S^n$ and, for spin 1/2, to general tangent frames. We worked out in particular the cases of polar coordinates and stereographic coordinates, and we determined the local Lorentz transformations which bring the tangent frames in their final orientation.
The cases for spin 3/2, spin 2, and antisymmetric tensor fields are treated the same way, but lack of space does not permit to include details of these derivations. 
With this information one can easily reproduce the figures and tables in KRN.

We also showed that the vector spherical harmonics in Cartesian coordinates in Jackson's book on Electromagnetism are the same as those we obtained after we transformed to polar coordinates. We explained that the derivatives in the expressions for the spinor harmonics in KRN are due to the local Lorentz rotations of the tangent frames.

One can obtain the spectra for any coset manifold $G/H$ from only group theory, without having to know the explicit form of the spherical harmonics. One begins with a coset representative $L(z)$. Then $L^{-1} d L$, where $d$ is the exterior derivative, lies in the Lie algebra, so $L^{-1} d L=e^a K_a+\om^i H_i$, where $K_a$ are the coset generators, $H_i$ the subgroup generators, $e^a$ the coset vielbein one-form, and $\om^i$ the subgroup connection one-form. Next define $Y=L^{-1}$, and use $L^{-1}d L=-d Y Y^{-1}$ to obtain $(d+\om^i H_i)Y=-e^a K_a Y$. Introducing $\p_a=e_a{}^\al \p_\al$ where $e_a{}^\al$ is the inverse of $e_\al{}^a$ in $e^a=d z^\al e_\al{}^a$, we obtain
\bea
D_a Y=-K_a Y, \quad D_a=\p_a+\om_a{}^i H_i
\eea
On symmetric coset manifolds 
coset generators $H_i$ span a subgroup of SO($n$) and 
$\om^i$ is the spin connection. (A reductive coset manifold satisfies $[H_i,K_a]\subset K$, and a symmetric coset manifold satisfies in addition $[K_a,K_b]\subset H$.) Now act again with $D_{b}$
\bea
D_b D_a Y=-[\om_b{}^i H_i, K_a]Y-K_a D_b Y
\eea
Moving the first term on the right-hand side to the left-hand side, it completes the derivative $D_b$ to a derivative ${\cal D}_b$ where ${\cal D}_b$ contains also a term with the spin connection which acts on the indices $a$ of $D_a$
\bea
{\cal D}_b D_a Y=K_a K_b Y
\eea
Finally take the trace to obtain $\Box Y=\sum_a (K_a)^2 Y$. By writing $\sum_a(K_a)^2$ as a sum over all generators of the group $G$ minus a sum over the generators of the subgroup $H$, one finds the eigenvalues of the Laplacian in terms of two quadratic Casimir generators
\bea
\Box Y=\Big( C_2(G)-C_2(H) \Big) Y
\eea
This approach is worked out in detail in the remarkable 3-volume textbook of Castellani, D'Auria and Fr\'e in reference~\cite{reviews}{}. Each spherical harmonic corresponds to a Young tableau of the group $G$. The degeneracy of the eigenvalues is the dimension of the Young tableau. The values of the Casimir operators and their degeneracy can for example be found in reference~\cite{pilch}{}. For convenience we add these formulas, and examples, as an appendix.

\vspace{20pt}

It is a pleasure to thank L.~Castellani, R.~D'Auria, P.~Fr\'e, K.~Pilch, L.~Romans and B.~Schellekens for discussions and email correspondence.

\appendix
\section{Dimensions of irreps of $SO(2n)$ and eigenvalues of $C_2$ for $SO(n)$}
\subsection*{Dimensions of irreducible representations of $SO(2n)$}
Most people are familiar with the factors over hooks rule for $SU(n)$, but here we need formulas for $SO(n)$. We restrict ourselves to $SO(2n)$, because we have $G = SO(6)$. We also have $H = SO(5)$ but the irreps of $SO(5)$ are very simple, so we do not need the formulas for $SO(2n+1)$.

The irreducible representations (irreps) of $SO(2n)$ for tensors are denoted by Young tableaux as usual,  but the spinor irreps are denoted by adding a dot in each box.  For example
\eqa
\begin{array}{cc}
\ydiagram{1,1} & \ytabl{\bullet \\ \bullet} \\
t^{\mu\nu} = - t^{\nu\mu} & t^{\mu\nu}_A =- t^{\nu\mu}_A
\end{array}
\eqae
where the (chiral or antichiral) spinor index runs from $A=1$ to $A=2^{n-1}$.  

We label the rows for $SO(2n)$ from bottom to top by the row number $r_i$ where $r_i  = 0, \ldots , n-1$.  The number of boxes in each row is $\l_i$, with $\l_1 \leq \l_2 \leq \cdots \leq \l_n$ from bottom to top.  Then we add the $\l_i$ to $r_i$ for tensor irreps, or $\l_i + \frac12$ to $r_i$ for spinor irreps. (The addition of a dot in each box corresponds to the addition of $+\half$ in each $R_i$.) This given a second sequence of row numbers
\bea
R_i &=& (0 + \l_1 , \ldots , n-1 + \l_n ) \;  \mbox{for tensor irreps.} \nn\\
    &=&  \left( 0 + \l_1 + \tfrac{1}{2} , \ldots , n-1 + \l_n + \tfrac{1}{2} \right) \; \mbox{for spinor irreps.}
\eea
The degeneracy formula (the dimension of the representation) is then
 \eqa
 d= {\Pi \; {\rm Sums} \; \Pi \; {\rm Differences} \over \Pi \; {\rm sums} \; \Pi \; {\rm differences}}
 \eqae
where Sums (Differences) are all $R_j + R_i \; (R_j - R_i)$, and sums (differences) are all $r_j + r_i \; (r_j - r_i)$, with always $j>i$.

We now give some examples for $SO(6)$.  Of course, $SO(6)$ is isomorphic to $SU(4)$, so one can also get these results from the ``factors over hooks" rule for unitary groups \cite{GeorgiH}.
\\\\
\noindent 1)  antisymmetric tensor 
 \eqa
\begin{array}{l} r_3 = 2 \\ r_2 =1 \\ r_1 =0 \end{array}
\begin{array}{c}
\boxup2{\boxes2} \\ \dots
\end{array}
\begin{array}{l}
\l_3 =1 \\ \l_2 =1 \\ \l_1 =0 \end{array}
\begin{array}{l} R_3 = 3  \\ R_2 = 2  \\ R_1 = 0  \end{array}
\begin{array}{l} {\rm Sums} \;  = (5,3,2) \; {\rm Differences} \;  = (1,3,2) \\ {\rm sums} \;  = (3,2,1) \; {\rm differences} \;  = (1,2,1)
\end{array}
\eqae
\eqa
d = {(5 \cdot 3 \cdot 2) \cdot (1 \cdot 3 \cdot 2) \over (3 \cdot 2 \cdot 1) \cdot ( 1 \cdot 2 \cdot 1)} = {\bf 15}
\eqae
Indeed, the tensor $t^{\mu\nu} =- t^{\nu\mu}$ has ${6 \choose 2} =15$ states. It is the adjoint representation of $SO(6)$, and corresponds to $\boxup3{\boxes3\boxes1}$ in $SU(4)$ which is also the adjoint representation.
\\\\
 \noindent 2) antisymmetric tensor-spinor
\eqa
\begin{array}{l} r_3 = 2 \\ r_2 =1 \\ r_1 =0 \end{array}
\begin{array}{l}
   \begin{tabular}{|l|ll}
\hline
$\bullet$   \\
\hline
$\bullet$ \\
\cline{1-1}
\end{tabular}
\end{array}
\begin{array}{l}
\l_3 =1 \\ \l_2 =1 \\ \l_1 =0 \end{array}
\begin{array}{l} R_3 = 3 + \frac12 \\ R_2 = 2 + \frac12 \\ R_1 = \frac12  \end{array}
\begin{array}{l} {\rm Sums} \;  = (6,4,3) \; {\rm Differences} \;  = (1,3,2) \\ {\rm sums} \;  = (3,2,1) \; {\rm differences} \;  = (1,2,1)
\end{array}
\eqae
\eqa
d = {(6 \cdot 4 \cdot 3) \cdot (1 \cdot 3 \cdot 2) \over (3 \cdot 2 \cdot 1) \cdot ( 1 \cdot 2 \cdot 1)} = {\bf 36}
\eqae
Indeed, the tensor-spinor $t^{\mu\nu}_A =- t^{\mu\nu}_A$ has ${6 \choose 2} \cdot 4 =60$ states, but the gamma-traceless irreducibility condition $\g_\mu t^{\mu\nu}_A =0$ subtracts $6 \times 4 =24$ states, leaving 36 states. In $SU(4)$, we obtain this irrep as follows
\begin{equation}
\begin{array}{ccccccccccccc}
 \boxup3{\boxes3\boxes1} &\otimes&  \boxup1{\boxes1} &=&  \boxup3{\boxes3\boxes1\boxes1} &\oplus& \boxup3{\boxes3\boxes2} &\oplus& \boxup4{\boxes4\boxes1}  \\
 {\bf 15} &\times& {\bf 4} &= & {\bf 36}  &+& {\bf 20} &+& {\bf 4}
\end{array}
\end{equation}
\subsection*{Eigenvalues of the quadratic Casimir operator for SO(n)}
We now discuss how to obtain the eigenvalues of the Casimir operators $C_2(G)$ and $C_2(H)$ both for tensorial and for spinorial representations of $SO(n)$. First tensors.

Consider a tensor $T_{\a_1 \cdots \a_r}$ in an irreducible representation (irrep) of $SO(n)$ with $r$ vector indices, with the symmetry of a given Young diagram.  Let $f_i$ and $g_j$ be the number of boxes in the $i-th$ row and $j-th$ column, respectively.  Then the eigenvalues of $C_2 [SO(n)]$ for tensors are given by \cite{pilch}
\eqa
C_2  [SO(n)]_{\rm tensors} \;  =- r (n-1) -2 \O
\eqae
where $r$ is the number of boxes and
\eqa
2 \O = \sum f^2_i - \sum g^2_j
\eqae
For example, consider the following irrep of $SO(6)$
\eqa
\begin{array}{l}
   \boxup3{\boxes3\boxes2\boxes1\boxes1}  \end{array}
\begin{array}{l}
f_1 =4 \\ f_2 =2 \\ f_3 =1 \end{array}
\begin{array}{l} g_1 =3  \\ g_2=2 \\ g_3 = g_4 =1  \end{array}
\begin{array}{l} r=7  \end{array}
\eqae
In this example, $2 \O$ is equal to 6, and the eigenvalue of the Laplacian for this spherical harmonic is $-7.5-6=-41$.

Consider now antisymmetric tensors on $S^n$ with $k$ indices.  We denote $D^\a D_\a$ on $S^n$ by $\bo_S$.  For scalars (the case $k=0$) the corresponding Young tableaux are
\eqa
\begin{array}{ll}
G: \boxup1{\boxes1\boxes1\boxes1\boxes1\boxes1} & f_1 = l ; g_1 = \cdots = g_l = 1;  r_G = l =0,1,2 \cdots \\ H: \quad \bullet & \mbox{all} \; f_i = g_j =0 ; r_H =0 \end{array}
\eqae
where the dot denotes the trivial representation.  The eigenvalues of the Laplacian on scalar harmonics on $S^5$ yield then the eigenvalues
\bea
- \l_s &=& C_{2,SO(6)} ( \boxup1{\boxes1\boxes1\boxes1\boxes1\boxes1} ) - C_{2, SO(5)} ( \cdot ) \nn\\
&=& - 5l - l^2 + l \; ; \; l=0 , 1 , 2 \cdots
\eea
For the irrep $\bo$ one has $f_1 = g_1 =1, f_2 = f_3 = g_2 = g_3 =0$, so $2 \O =0$, and $C_2 (\bo) =-5$.  This is indeed the eigenvalue for a scalar on $S^5$ with $l=1$.

For transversal vectors (the case $k=1$), the $SO(6)$ Young tableaux are
\eqa
\begin{array}{l}
   \boxup2{\boxes2\boxes1\boxes1\boxes1\boxes1}  \end{array}
\begin{array}{l}
f_1 =l \\ f_2 =1 \\  r = l +1 \end{array}
\begin{array}{ll} g_1 =2 &  2 \O = l^2 + 1 - 4 - (l-1)    \\ g_2 = \cdots g_l =1  \end{array}
\eqae
while the $SO(5)$ representation of a vector corresponds to
\eqa
\bo  \qquad f_1 = 1 ; g_1 =1; r=1 ; 2 \O =0
\eqae
The eigenvalues of the Laplacian on transversal vector harmonics on $S^n$ are then
\eqa
- \l_v &=& [ - (l+1) 5- \{ l^2 -l-2 \} ] - [-4] \nn\\
&=& - [ l (l+4) -1 ]  \; ; \; l =  1 , 2 \cdots
\eqae

We turn now to antisymmetric tensors.  A two-index antisymmetric tensor corresponds to the following Young tableaux
\begin{eqnarray*}
\begin{array}{ll}
 G:  &   \boxup3{\boxes3\boxes1\boxes1\boxes1\boxes1} \\  & g_1 = 3 ; g_2 = \cdots g_l  = 1   \\ \end{array}
\begin{array}{l}
 f_1 =l \\ f_2 =1 \\  f_3 =1 \end{array}
\begin{array}{l} r = l+2 \\ 2 \O = l^2 + 2 - 9 - (l-1) \\ l=1,2, \cdots   \end{array}
\end{eqnarray*}
\eqa
\begin{array}{ll}
 H:  &   \boxup2{\boxes2} \\  & g_1 = 2  \end{array}
 \begin{array}{l} f_1 =1 \\ f_2 =1 \end{array}
\begin{array}{l}
r=2 \\ 2 \O  =- 2  \end{array} \hphantom{xxxxxxxxxxxxxxxxxxxxxxxxx}
\eqae
The eigenvalues of the Laplacian on $S^5$ acting on transversal antisymmetric tensors $Y_{\al\beta} =- Y_{\beta\al} , D^{\al} Y_{\al\beta} =0$ are thus given by
\bea
- \l (Y_{\al\beta}) &=&  [ - r (5) - 2 \O ]_G - [ -r (4) -2 \O ]_H \nn\\
&=&  [ - (l+2) 5- ( l^2 -l-6) ] + [  2 (4) - 2 ] \nn\\
&=& - [ l^2 + 4 l- 2 ] \; ; \; l= 1, 2, \cdots
\eea

We could consider next antisymmetric transversal tensors on $S^5$ with 3 indices, but since we do not need them, we stop here.\footnote{As a check one may consider the duals $\tilde{A}^a = \epsilon^{abc} D_b A_c$ in $d=3$, $\tilde{A}^{ab} = \epsilon^{abcd} D_c A_d$ in $d=4$, $\tilde{A}^{abc} = \epsilon^{abcde} D_d A_e$ in $d=5$, or $\tilde{A}^{a} = \epsilon^{abcd} D_b A_{cd}$ in $d=4$, etc. They are transversal due to the cyclic identity of the Riemann tensor, and give relations between the eigenvalues of different harmonics.} We just quote the result: $-\lambda (Y_{\al \beta \gamma})= -l^2 - 4l +3$. Finally we consider spinors. We just quote the formula for the eigenvalues of $C_2 (SO(n))$ on spinor harmonics.  (One can derive these results for spinors using coset methods.)  For Young diagrams with $r$ boxes (and a dot in each box to indicate that we are dealing with spinor harmonics) one finds \cite{pilch}
\eqa
C_2 [SO(n)]_{\rm spinors} =- \left[ rn + \tfrac{1}{8} n (n-1) +2 \O \right]
\eqae
For example, for spin $\frac{1}{2}$ spinors on $S^5$ one has
\begin{align}
G&: \ytabl{\bullet & \bullet & \bullet & \bullet & \bullet} \quad f_1 = l;\; g_1 = \cdots = g_l =1;\; 2 \O = l^2 -l;\; r=l = 0,1,2, \cdots \nn\\
H&: \bullet \quad f_i = g_j =0;\; 2 \O =0;\; r=0\,.
\end{align}
The dot indicates here the spinor representation without vector indices (either the fundamental $s$ or $c$ representation.  The eigenvalues are the same for $s$ and $c$).  One finds then on $S^5$
\bea
- \l_{\text{spin }\frac{1}{2}} &=& - \left[ l \cdot 6 + \tfrac{1}{8} \cdot 6 \cdot 5 + l^2 - l \right] + \left[\tfrac{1}{8} \cdot 5 \cdot 4 \right] \nn\\
&=& - \left( l^2 + 5 l + \tfrac{10}{8} \right) \; ; \; l =0, 1, 2 \cdots
\label{kk46}
\eea
The eigenvalues of the Dirac operator on $S^n$ with radius $1/c$ for spin $\frac12$ harmonics are $\pm i c \left( l + {n \over 2} \right)$.  Using
\eqa
\bo_S = \slashed{D} \slashed{D} - \tfrac{1}{4} R \; ; \; R= - 20 c^2 \; {\rm on} \;  S^5
\eqae
we find at unit radius $\bo_S =-  \left( l + {n \over 2} \right)^2 + 5$ with $n=5$.  This agrees with the result in (\ref{kk46}).


\begin{thebibliography}{99}
\bibitem{KRN} 
H.~J.~Kim, L.~J.~Romans and P.~van~Nieuwenhuizen, {\it Phys. Rev. D} {\bf 32} (1985) 389.
\bibitem{reviews} 
For reviews of spherical harmonics see \\A. Salam and J. Strathdee, {\it Ann.\ Phys.} {\bf 141} (1982) 316 (a classic). \\ R. D'Auria and P. Fr\'e, {\it Ann.\ Phys.} {\bf 157} (1984) 1 and {\bf 162} (1985) 372. \\ P. van Nieuwenhuizen, lectures at Les Houches 1983 and Trieste school 1984.\\ M.J. Duff, B.E.W. Nilsson and C.N. Pope, {\it Phys.\ Rep.} {\bf 130} (1986) 1. \\ L. Castellani, R. D'Auria and P. Fr\'e, ``{\it Supergravity and Superstrings, a Geometric Perspective}", vol. 2, ch. 5, World Scientific 1991.
\bibitem{kreuzer}
Max Kreuzer at http://hep.itp.tuwien.ac.at/$\sim$kreuzer/.
\bibitem{chodos} 
A.~Chodos and E.~Myers, {\it Ann. Phys.} {\bf 156} (1984) 412.
\bibitem{rubin} 
M.~Rubin and C.~R.~Ordonez, {\it J. Math. Phys.} {\bf 25} (1984) 2888.
\bibitem{Jackson}
J.~D.~Jackson,  {\it ``Classical Electrodynamics"}, 2nd edition, Willey 1975, section 16.2.
\bibitem{gibbons} 
G.~W.~Gibbons and A.~R.~Steif, {\it Phys. Lett. B} {\bf 320} (1994) 245.
\bibitem{pilch} 
K.~Pilch and A.~N.~Schellekens, {\it J. Math. Phys.} {\bf 25} (1984) 3455.
\bibitem{GeorgiH} H. Georgi, ``{\it Lie algebras in particle physics}", Benjamin/Cummings 1982, chapter XIII.
\end{thebibliography}
\end{document}